\documentclass[preprint2]{aastex63}

\newcommand{\zaphot}{0.63 \pm 0.12}
\newcommand{\zbphot}{0.60 \pm 0.17}
\newcommand{\rf}{\emph{realfast}}
\usepackage{color}
\newcommand{\add}[1]{{#1}}  
\newcommand\remove[1]{\unskip}  

\newcommand{\frbname}{FRB 20190614D}
\newcommand{\dmunits}{{\rm pc\,cm^{-3}}}
\newcommand{\eg}{e.\,g.}

\graphicspath{{./}{figures/}}
\begin{document}

\title{A Distant Fast Radio Burst Associated to its Host Galaxy with the Very Large Array}

\correspondingauthor{Casey J. Law}
\email{claw@astro.caltech.edu}

\author[0000-0002-4119-9963]{Casey J. Law}
\affiliation{Cahill Center for Astronomy and Astrophysics, MC 249-17 California Institute of Technology, Pasadena, CA 91125, USA}

\author[0000-0002-5344-820X]{Bryan J. Butler}
\affiliation{National Radio Astronomy Observatory, Socorro, NM, 87801, USA}

\author{J. Xavier Prochaska}
\affiliation{University of California Observatories-Lick Observatory, University of California, 1156 High Street, Santa Cruz, CA 95064, USA}
\affiliation{Kavli Institute for the Physics and Mathematics of the Universe, 5-1-5 Kashiwanoha, Kashiwa, 277-8583, Japan}

\author[0000-0001-5162-9501]{Barak Zackay}
\affiliation{Institute for Advanced Study, Princeton}

\author[0000-0003-4052-7838]{Sarah Burke-Spolaor}
\affiliation{Center for Gravitational Waves and Cosmology, West Virginia University, Chestnut Ridge Research Building, Morgantown, WV 26505}
\affiliation{Department of Physics and Astronomy, West Virginia University, Morgantown, WV 26506}
\affiliation{CIFAR Azrieli Global Scholars program, CIFAR, Toronto, Canada}

\author{Alexandra Mannings}
\affiliation{University of California Observatories-Lick Observatory, University of California, 1156 High Street, Santa Cruz, CA 95064, USA}

\author[0000-0002-1883-4252]{Nicolas Tejos}
\affiliation{Instituto de F\'isica, Pontificia Universidad Cat\'olica de Valpara\'iso, Casilla 4059, Valpara\'iso, Chile}

\author[0000-0003-3059-6223]{Alexander Josephy}
\affiliation{Department of Physics, McGill University, 3600 University Street, Montr\'{e}al, QC H3A 2T8, Canada}
\affiliation{McGill Space Institute, McGill University, 3550 University Street, Montr\'{e}al, QC H3A 2A7, Canada}

\author[0000-0001-5908-3152]{Bridget Andersen}
\affiliation{Department of Physics, McGill University, 3600 University Street, Montr\'{e}al, QC H3A 2T8, Canada}
\affiliation{McGill Space Institute, McGill University, 3550 University Street, Montr\'{e}al, QC H3A 2A7, Canada}

\author[0000-0002-3426-7606]{Pragya Chawla}
\affiliation{Department of Physics, McGill University, 3600 University Street, Montr\'{e}al, QC H3A 2T8, Canada}
\affiliation{McGill Space Institute, McGill University, 3550 University Street, Montr\'{e}al, QC H3A 2A7, Canada}

\author[0000-0002-9389-7413]{Kasper E. Heintz}
\affiliation{Centre for Astrophysics and Cosmology, Science Institute, University of Iceland, Dunhagi 5, 107 Reykjav\`ik, Iceland}

\author[0000-0002-2059-0525]{Kshitij Aggarwal}
\affiliation{Department of Physics and Astronomy, West Virginia University, Morgantown, WV 26506}

\author[0000-0003-4056-9982]{Geoffrey C. Bower}
\affiliation{Academia Sinica Institute of Astronomy and Astrophysics, 645 N. A'ohoku Place, Hilo, HI 96720, USA}

\author[0000-0002-6664-965X]{Paul B. Demorest}
\affiliation{National Radio Astronomy Observatory, Socorro, NM, 87801, USA}

\author{Charles D. Kilpatrick} 
\affiliation{University of California Observatories-Lick Observatory, University of California, 1156 High Street, Santa Cruz, CA 95064, USA}

\author{T. Joseph W. Lazio}
\affiliation{Jet Propulsion Laboratory, California Institute of Technology, 4800 Oak Grove Dr, M/S 67-201, Pasadena, CA 91109 USA}

\author[0000-0002-3873-5497]{Justin Linford}
\affiliation{National Radio Astronomy Observatory, Socorro, NM, 87801, USA}

\author{Ryan Mckinven}
\affiliation{Department of Astronomy and Astrophysics, University of Toronto, 50 St. George Street, Toronto, ON M5S 3H4, Canada}
\affiliation{Dunlap Institute for Astronomy and Astrophysics, University of Toronto, 50 St. George Street, Toronto, ON M5S 3H4, Canada}

\author[0000-0003-2548-2926]{Shriharsh Tendulkar}
\affiliation{Department of Physics, McGill University, 3600 University Street, Montr\'{e}al, QC H3A 2T8, Canada}
\affiliation{McGill Space Institute, McGill University, 3550 University Street, Montr\'{e}al, QC H3A 2A7, Canada}

\author[0000-0003-3801-1496]{Sunil Simha}
\affiliation{University of California Observatories-Lick Observatory, University of California, 1156 High Street, Santa Cruz, CA 95064, USA}


\begin{abstract}
We present the discovery and subarcsecond localization of a new Fast Radio Burst with the Karl G. Jansky Very Large Array and \rf\ search system. The FRB was discovered on 2019 June~14 with a dispersion measure of 959~$\dmunits$. This is the highest DM of any localized FRB and its measured burst fluence of 0.6~Jy~ms is less than nearly all other FRBs. The source is not detected to repeat in 15 hours of VLA observing and 153 hours of CHIME/FRB observing. We describe a suite of statistical and data quality tests we used to verify the significance of the event and its localization precision. Follow-up optical/infrared photometry with Keck and Gemini associate the FRB to a pair of galaxies with $\rm{r}\sim23$\ mag. The false-alarm rate for radio transients of this significance that are associated with a host galaxy is roughly $3\times10^{-4}\ \rm{hr}^{-1}$. The two putative host galaxies have similar photometric redshifts of $z_{\rm{phot}}\sim0.6$, but different colors and stellar masses. Comparing the host distance to that implied by the dispersion measure suggests a modest ($\sim 50~\dmunits$) electron column density associated with the FRB environment or host galaxy/galaxies.
\end{abstract}

\section{Introduction}

Fast Radio Bursts (FRBs) are millisecond-timescale radio transients of extremely high brightness originating at cosmological distances \citep{2019A&ARv..27....4P,2019ARA&A..57..417C}. More than hundreds of FRBs are known currently, and the inferred occurrence rate is roughly $10^3$~sky$^{-1}$~day$^{-1}$ above a fluence limit of 1~Jy~ms at frequencies near 1.4~GHz \citep{2016MNRAS.460L..30C,2017AJ....154..117L}. FRB distances can be estimated from the dispersive delay induced by propagation through ionized gas (quantified by a Dispersion Measure, \hbox{DM}, which measures the total electron column density along the line of sight to the source); for FRBs, the measured DMs are significantly larger than those expected due to contributions from our own Galaxy \citep{2002astro.ph..7156C}. By attributing the dispersion induced outside of our Galaxy to predictions for the intergalactic medium (IGM), FRBs are estimated to originate at characteristic distances one to a few gigaparsecs \citep{2004MNRAS.348..999I,2007Sci...318..777L}. Several FRBs have been localized by radio interferometers and associated with host galaxies of known distance; their luminosity distances range from 149 Mpc to 4 Gpc \citep{2020Natur.577..190M,OPT,2019Sci...365..565B, 2019Sci...366..231P,2019Natur.572..352R,2020arXiv200513161M}.

It is not yet known what causes FRBs or whether there are multiple formation channels \citep{2016MNRAS.461L.122L,2019NatAs...3..928R}. Identifications of FRB host galaxies is a critical test of formation models, as it can constrain the age of the stellar populations in FRB environments. The first host galaxy suggested that FRBs are associated with peculiar star-forming environments \citep{2017ApJ...843L...8B} but later hosts have a wider range of environments \citep{2019Sci...365..565B,2019Natur.572..352R,2020arXiv200513160B}.

Radio waves are modified as they propagate through ionized gas \citep[e.g., dispersion, scattering, lensing, Faraday rotation;][]{2017ApJ...842...35C,2019MNRAS.485L..78V}. This fact, combined with the large distance to FRBs, makes them novel probes of the IGM and other galaxies \citep{1973Natur.246..415G,2015Natur.528..523M,2019Sci...366..231P}. Furthermore, the fact that dispersion is an unambiguous tracer of baryonic mass has opened potential for FRBs to study galaxy halos and cosmology \citep{2019ApJ...872...88R,prochaska+zheng2019}. However, most of this science potential can only be achieved by measuring distances to FRBs.  Multiple radio interferometers for precision FRB localization are in phases of conceptual development, construction, or commissioning \citep{2018ApJS..236....8L,2019MNRAS.489..919K,2019Sci...365..565B,2019ATel13098....1C,2017ATel10693....1O}. The goal of all these projects is to localize FRBs to arcsecond precision, which is required to unambiguously associate it to a host galaxy \citep{2018ApJ...860...73E}.

Many FRBs are seemingly single flashes, and before the advent of widespread of use of GPUs to accelerate complex processing, single-dish telescopes generally led blind searches for new FRBs \citep{2011MNRAS.416.2465B,2013Sci...341...53T,2014ApJ...790..101S}.
However, some FRBs, such as FRB 121102, emit multiple bursts at irregular intervals \citep{2016Natur.531..202S,2018ApJ...866..149Z}, which made it possible to target with interferometers \citep{2017Natur.541...58C, EVN}. The Canadian Hydrogen Intensity Mapping Experiment (CHIME) is a transit telescope operating between 400--800 MHz that is rapidly discovering both repeating and non-repeating FRBs \citep{2018ApJ...863...48C,2019arXiv190803507T}. The CHIME/FRB search has a localization precision of roughly 10\arcmin, which is too large to unambiguously identify host galaxies for FRBs.

Here, we present a new FRB discovery and localization by the Karl G. Jansky Very Large Array (VLA) using \rf\  \citep{2018ApJS..236....8L}. The FRB was found coincidentally during a search for CHIME/FRB FRB~180814.J0422$+$73 \citep[hereafter FRB~180814,][]{2019Natur.566..235C}. This new FRB is associated with a unique host galaxy with a distance that is consistent with expectations for its \hbox{DM}. The combination of radio interferometric data and optical associations support the conclusion that it is a new FRB and we refer to it as \frbname. We discuss the FRB environment and constraints on the  distribution of DM in the IGM and host galaxy.

\section{Observations}\label{sec:observe}

\subsection{Program and Overall Description}\label{sec:program}

In 2018, the VLA and CHIME/FRB teams began collaborating to use the VLA for follow-up of repeating FRBs found by CHIME/FRB. We have had two approved projects: VLA/18B-405 and VLA/19A-331.
We targeted FRB180916.J0158+65 and FRB190303.J1353+48 for 40 hours scheduled under VLA/18B-405 and FRB~180814 for 39 hours scheduled under VLA/19A-331; this paper focuses on the second project.

We observed using the L-band system of the VLA, spanning 1-2 GHz \remove{with spectral channels of 1~MHz width}, in twenty separate observations. We observed a field centered at (\hbox{RA}, Dec) [J2000] = (04h22m22s, $+73$d40m00s), the approximate position of FRB~180814. The nominal field of view of the VLA antennas at L-band is $\sim\!30\arcmin$ (full width half maximum at 1.4~GHz), but the \rf\ system is configured to image a field 2 times wider than that.
The first seven observations were performed in December 2018, in the C-configuration of the VLA, with maximum baselines $\sim\!3$~km and a resolution of $\sim\!14\arcsec$ at 1.4 GHz.  
Thirteen later observations were performed in February through July of 2019, in the B- or BnA-configurations of the VLA, with maximum baselines $\sim\!11$~km in length and a resolution of $\sim\!4.5\arcsec$ at 1.4 GHz. Each observation 
had an on-source time of around 1.5 hours
that was searched by the \rf\ system.
The detection reported here is the strongest FRB-like event found in this campaign and is the focus of the analysis presented.

\subsection{Search Technique}\label{sec:search}

The observations used a commensal correlator mode that generated visibilities with an integration time of 5~ms to be searched by \rf. The same data also were used to generate and save the standard visibility data product to the NRAO archive with a sampling time of 3~s, for all observations in June and July 2019 (nine of them). Prior to that, all visibilities were saved to the archive at their full time resolution, resulting in large datasets (of order 1.5 TB). \add{Both fast and slow visibilities were made in 16 64-channel spectral windows, with each channel set to a width of 1~MHz. Taking typical interference flagging into account, the usable bandwidth is 600~MHz.}

The fast-sampled visibilities were distributed to a dedicated GPU cluster using \texttt{vysmaw} \citep{2018JAI.....750005P} and searched with \texttt{rfpipe} \citep{2017ascl.soft10002L}. After applying available on-line calibrations, the search pipeline dedispersed and integrated visibilities in time before forming images. \add{Calibration solutions derived from $\sim$minute-long scans and are stable in time (less than 5 deg change from mean value).} Images were generated with a simple, custom algorithm that uses natural weighting and a pillbox gridding scheme. The search used 215 DM values from~0 to 1000~$\dmunits$\ and four temporal widths from~5 to 40~ms, which is inclusive of the known properties of \add{FRB180814} \remove{FRB180916} \citep{2019Natur.566..235C}.


For the B-configuration observations, each image had 2048$\times$2048 pixels with a pixel size of 1.7\arcsec, covering a field of view of 1\arcdeg. The C-configuration images were 512$\times$512 pixels with a pixel size of roughly 6.8\arcsec. The nominal 1$\sigma$\ sensitivity in a single 5~ms integration \remove{with 600~MHz bandwidth (taking into account typical RFI flagging)} is 6~mJy~beam$^{-1}$. All candidates detected with significance greater than 7.5$\sigma$\ trigger the recording of 2--3~s of fast sampled visibilities and a visualization of the candidate. Each candidate is classified by \texttt{fetch}, a convolutional neural network for radio transients \citep{2019arXiv190206343A}. Finally, \rf\ team members review the visualizations of the real-time analysis to either remove data corrupted by interference or identify candidates for more refined offline analysis.

\subsection{Discovery}\label{sec:discovery}

On~2019 June~14 (UT), the \rf\ system detected a candidate transient in the FRB~180814 field. The realtime detection system reported a candidate with image significance of 8.0$\sigma$\ and $\mathrm{DM} = 959\,\dmunits$, far in excess of the expected DM contribution of the Milky Way \citep[83.5~$\dmunits$,][]{2002astro.ph..7156C}.  However, the DM of FRB~180814 is 189.4~pc~cm$^{-3}$; no FRB has shown changes in DM of more than a few pc~cm$^{-3}$ \citep{2018ApJ...863....2G}, so the candidate FRB is likely unrelated to the CHIME \hbox{FRB}.

\begin{figure}[tb]
\begin{center}
\includegraphics[width=\columnwidth,trim=5mm 10mm 20mm 15mm,clip]{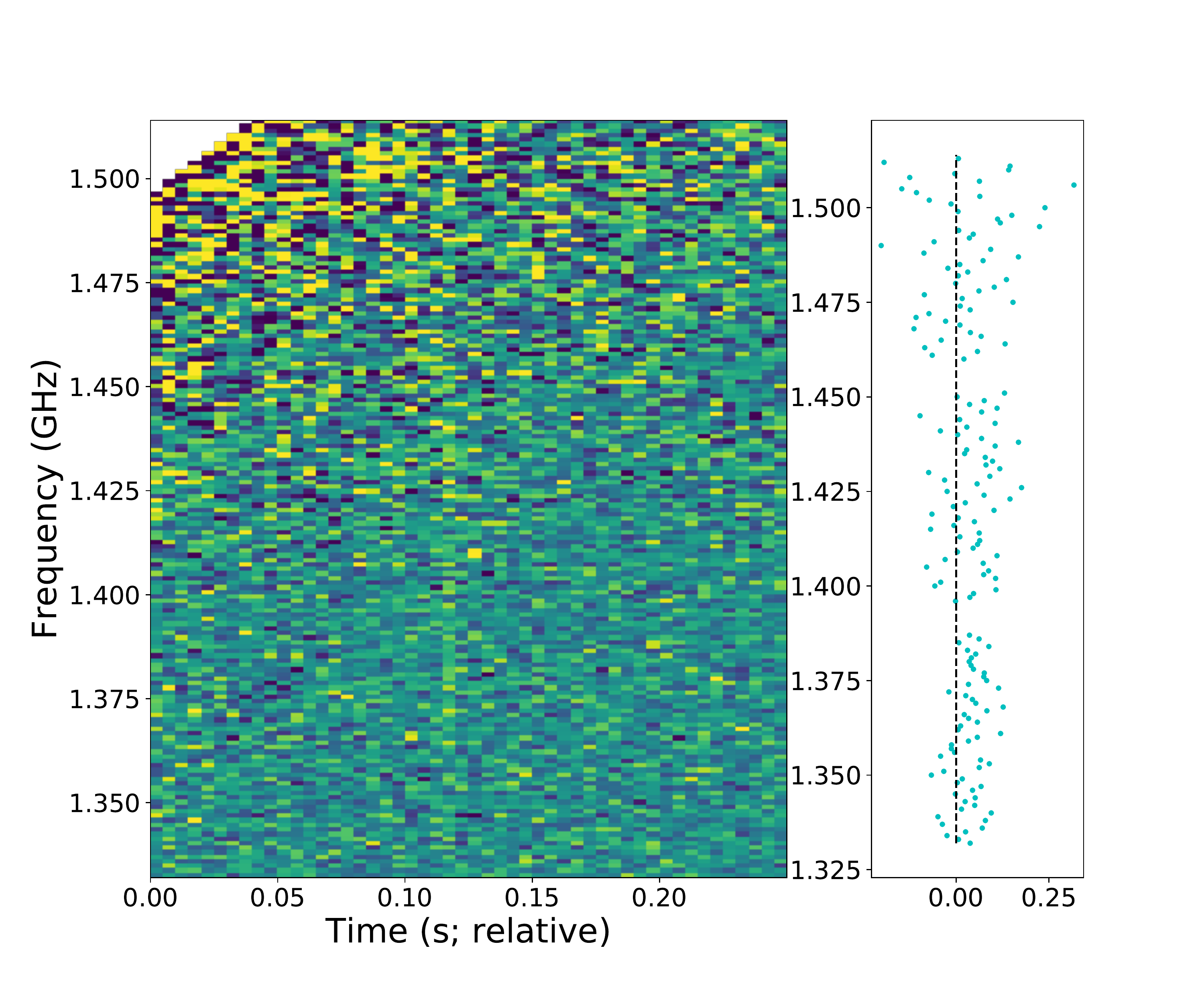}
\caption{(Left) Stokes~I dynamic spectrum for the candidate FRB as seen by VLA/\rf. The dynamic spectrum was generated by summing calibrated visibilities for all baselines and the two orthogonal polarizations. The gap and higher noise level toward the top left of the dynamic spectrum results from when the data recording was initiated.
(Right) Stokes~I spectrum taken from a single 5~ms integration of the dynamic spectrum.
\label{fig:dynspec}}
\end{center}
\end{figure}

The realtime candidate analysis revealed multiple signatures consistent with an astrophysical source. First, the spectrum (Figure~\ref{fig:dynspec}, right panel) shows emission over a range of frequencies spanning at least 50~MHz and the image shows a compact source. Most sources of interference tend to have circular polarization, narrow spectral extent, or are spatially incoherent (i.e., radio frequency interference in the near-field of the array). 
Second, the \texttt{fetch} FRB classification system reported an astrophysical probability of 99.9\%. Third, there is a weak prior expectation for blindly-detected astrophysical events to be detected where the antenna sensitivity is highest. The candidate was detected roughly 9\arcmin\ away from the pointing center, where the antenna has roughly 80\% of its nominal sensitivity; only 10\% of the image has this sensitivity or higher.

The \rf\ search system was starting to receive visibilities from the VLA correlator during the burst. This is seen in Figure \ref{fig:dynspec}, which shows that the mean of all recorded visibilities during the burst (phased toward the event) is noisier at early times and at higher frequencies. Visibilities for each baseline, polarization, and spectral window (64 channels) are distributed separately such that the fraction of data grows to 100\% over a few hundred milliseconds as the system turns on.

\subsection{Verification Tests and Significance Analysis}\label{sec:verify}

Traditional fast transient surveys measure event significance based on a noise estimate that is local in time (e.g., a standard deviation of a time series). Our interferometric search measures significance in a single image, so the noise estimate is made simultaneously. Appendix \ref{sec:std} describes how the visibility domain search can be thought of as a time-domain search that allows for more accurate noise estimates.

In our initial analysis of the candidate, we confirmed that the event significance was not affected by different flagging algorithms or calibration solutions from a calibrator observation \add{a few minutes after the event}. We also confirmed that removing an antenna from the 27-antenna array reduced the detection significance by roughly 5\% ($\approx 1/27$ antennas). With confidence in the quality of data, we proceeded to more carefully quantify the event significance.

We used the raw, saved visibilities to re-run the search with a larger image (8192$\times$8192 pixels) and finer DM grid. This optimized search improved the detection significance slightly to a S/N ratio of~8.27 at $\mathrm{DM} = 959.19$~$\dmunits$. Using the same refinement procedure on other candidates typically does not reproduce the initial detection. Noise-like events are expected to be sensitive to the image gridding parameters, so we ignore all events that cannot be reproduced in larger images. We use these refined properties for visualizations and all further analysis.

\begin{figure}[tb]
\begin{center}
\includegraphics[width=\columnwidth]{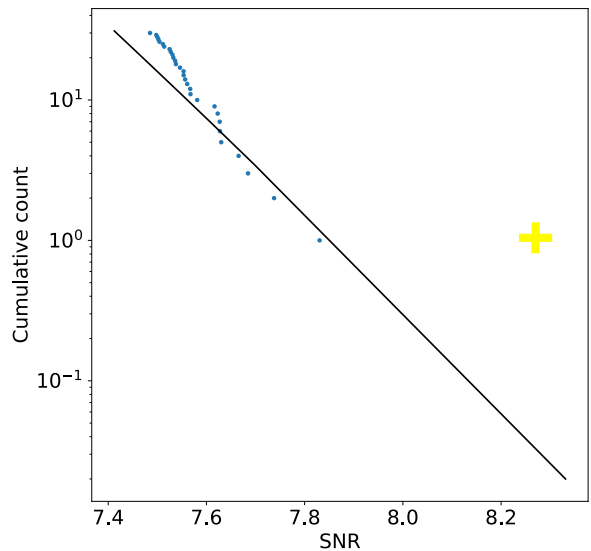}
\caption{Circles show the cumulative distribution of candidates in this observing campaign as a function of image S/N ratio. The solid line shows the expected cumulative event rate for a Gaussian (noise-like) S/N distribution. The yellow cross shows the candidate FRB S/N ratio after refinement analysis.
\label{fig:snrdist}}
\end{center}
\end{figure}

Figure \ref{fig:snrdist} shows the cumulative distribution of event significance for all events seen in this campaign. The FRB search pipeline automatically applies flags for bad calibration, antenna state, missing data, and interference. We visually inspected the 263 candidates detected above 7.5$\sigma$ in observations of this field and removed those affected by unflagged interference to get a sample of 31 candidates.

Figure \ref{fig:snrdist} also shows an \add{independent} estimate of the ideal event rate significance distribution for the array and correlator configuration used to find this candidate. The ideal cumulative event rate assumes that each pixel imaged has a brightness that is drawn from a stationary Gaussian distribution. The number of independent pixels searched is $(\rm{N}_{pix}/O_{pix})^2 \times \rm{N}_{int} * (\rm{N}_{DM}/O_{DM})$, where N$_{\rm{pix}}$ is the width of an image in pixels, N$_{\rm{int}}$ number of integrations (at all time widths), N$_{\rm{DM}}$\ is the number of DM trials, and O$_{\rm{pix}/\rm{DM}}$ are the oversampling of the synthesized beam and dispersion sensitivity function, respectively. Both images and DMs are oversampled to maintain uniform sensitivity to all locations and DMs. The search run here uses O$_{\rm{pix}}=2.5$ and O$_{\rm{DM}}=3$. In this configuration, we have 8.4~hrs of observing time and $5\times10^{14}$\ independent pixels. The candidate signal-to-noise ratio (S/N) of 8.27 corresponds to a False Alarm Rate (FAR) of once in 250~hr. The measured and ideal distributions are \add{independent and} in rough agreement, which shows that the \add{significance follows a Gaussian distribution and that this} candidate is an outlier.

%

The FRB search pipeline also uses spectral brightness fluctuations to distinguish candidate events from noise \citep{2017ApJ...850...76L,2019arXiv190803507T}. The Kalman detector (Zackay, in prep) is a method to estimate the statistical significance of FRB spectral variations for an assumed noise model and signal smoothness. For a given noise and signal model, we can marginalize the detection statistic over all matched filters, weighted by their prior probability. This prior probability is defined by a random walk with one free parameter, the coherence bandwidth. We calculated the Kalman score on the candidate FRB, using logarithmic spaced options for the smoothing scale, but found no significant change in the total confidence for the candidate FRB (other FRBs do show some improvement; Zackay, in prep). We conclude that the candidate FRB spectrum is consistent with a constant flux density.

\subsection{Localization}\label{sec:locate}

The realtime FRB search software makes several assumptions to improve computational efficiency, and as a result images which are used within it are not optimal. To address this, we used the stored raw, de-dispersed visibilities to re-image the burst data with a combination of CASA \citep{2007ASPC..376..127M} and AIPS \citep{2003ASSL..285..109G}\footnote{Both CASA and AIPS calibrate with a different algorithm from that used by the realtime calibration system known as ``telcal'' \citep{2018ApJS..236....8L}.}. Here, we describe a unified calibration and imaging procedure used in both fast and deep imaging. This procedure allows us to quantify the systematic error in the FRB localization.

Prior to re-imaging the burst data, we reduced all of the data taken in June and July 2019 for a deep image of the field. Nine datasets during B configuration were included in this analysis. We excluded C configuration data, as it has poorer spatial resolution. We also excluded early B configuration data recorded at the fast sampling rate, as it was computationally expensive to include in the deep imaging analysis.

We started by applying the calibration and flagging tables for each observation which were provided by the VLA calibration pipeline. For all observations, the flux density scale was set with an observation of the calibrator source 3C~147, and at these frequencies is accurate to 1-2\% \citep{2017ApJS..230....7P}. Bandpass and delay calibrations were also determined by the 3C~147 observation. Complex gain (amplitude and phase) fluctuations over time were calibrated with observations of the calibrator source J0410+7656 every 30 minutes. We then exported the calibrated visibilities from CASA and imported them into AIPS. After further RFI flagging, we averaged in time (to 9 seconds), and frequency (to 4 MHz channels) to reduce the computational load for the imaging. 

We used faceted imaging in AIPS to image to beyond the first null of the antenna primary beam response (1.1\arcdeg\ width). A total of 73 separate fields, each $1024\times1024$~ pixels (with 0.5\arcsec\ pixel size), and 250 CLEAN boxes were used to image and clean the area.  
After cleaning, the 73 images of the fields were combined together, and that result was used to self-calibrate \citep{1999ASPC..180..187C} the visibilities on a 1-minute timescale. The imaging and self-calibration was then repeated using this self-calibrated dataset, on a 9-second timescale - essentially self-calibrating every visibility.  
A final image was then made, and a primary beam correction made to it, based on \citet{Perley_beam_memo}. This is the final deep image used for further analysis. The synthesized beam in this final deep image is 3.6\arcsec\ $\times$ 2.8\arcsec\ at a position angle of 79\arcdeg\ (North through East).  The image has a $1\sigma$ sensitivity of ~3.6~$\mu$Jy~beam${}^{-1}$, consistent with expectations for the total on-source time and flagging.

For the re-imaging of the burst data, we first copied the VLA calibration pipeline tables (calibration and flagging) from the full June 14th observation, and ran a modified version of the procedure to re-apply these tables. 
Calibration tables from the three spectral windows (384 MHz of bandwidth) with valid, uncorrupted data were applied.
The synthesized beam in this final burst image is 10.3\arcsec\ $\times$ 4.2\arcsec\ at a position angle of 67\arcdeg. It is significantly worse than the resolution of the deep image because of the drastically reduced amount of data that went into it.

The deep and fast radio images were exported to CASA format for source detection and modeling.  The source detected by the \rf\ system (using \texttt{rfpipe}) is also detected in the burst image. We fit an ellipse to that source to measure the centroid location, peak flux density, and their 1$\sigma$\ uncertainties (see Table \ref{tab:prop}).
The localization precision is approximately 1/$10^{\mathrm{th}}$ of the synthesized beam diameter, which is typical for sources of this significance observed with the VLA \citep{1995ApJ...450..559B}.

\begin{figure}[tb]
\begin{center}
\includegraphics[width=\columnwidth]{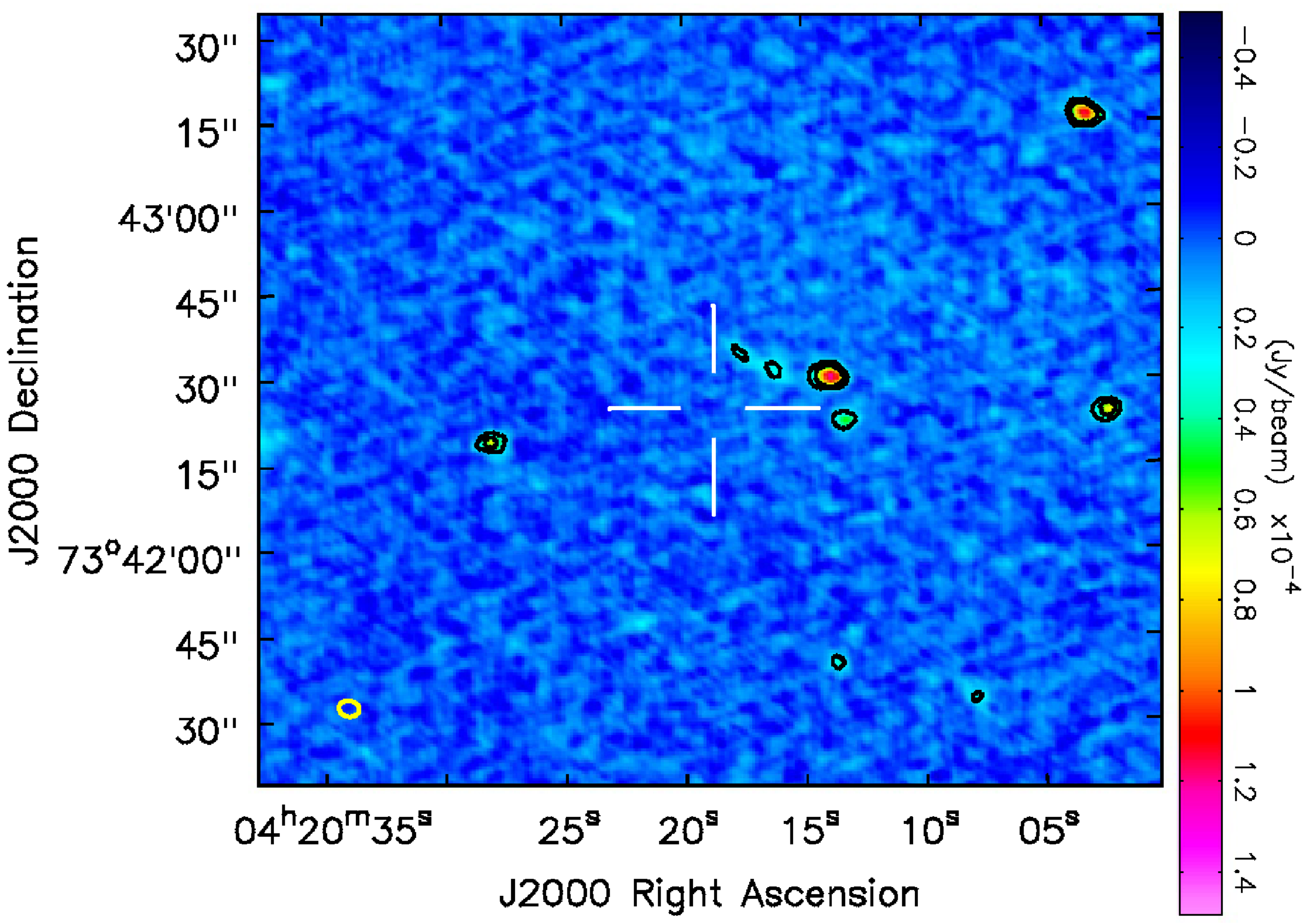}
\caption{Deep 1.4~GHz radio image of the FRB~180814 field with the location of \frbname\ shown with white cross-hairs. Black contours show radio brightness levels of 25 and 50~$\mu$Jy. No persistent radio emission brighter than 3$\sigma$~ (11 $\mu$Jy) is seen at the location of the new \hbox{FRB}. The noise level of this image is 3.6~$\mu$Jy~beam${}^{-1}$, and the beam shape is (3.6\arcsec, 2.8\arcsec, 78\arcdeg), marked by a yellow ellipse in the bottom left corner of the image.
\label{fig:imager}}
\end{center}
\end{figure}

We then searched the deep image to determine whether there is persistent radio emission associated with the candidate FRB. We find no such associated persistent radio emission at the location of the candidate FRB, to a 3$\sigma$\ limit of 11~$\mu$Jy (see Figure~\ref{fig:imager}).

We tested the astrometric precision by associating compact radio sources with optical sources in Pan-STARRS DR2 catalog \citep{2016arXiv161205560C}. We ran the \texttt{aegean} source finding package \citep{2018PASA...35...11H} and identified 270 compact radio sources with a flux density greater than 100$\mu$Jy ($>25\sigma$). Of these, 102 had optical counterparts within 3\arcsec\ and $\rm{nDetections}=5$. No systematic offset is found between the radio and optical sources; the standard deviation of the radio/optical offsets is 0.2\arcsec. We note that given the resolution of the radio image (3.6\arcsec\ $\times$ 2.8\arcsec), we expect the astrometric accuracy to be of the order of 0.1\arcsec\ for these brighter sources (a few \% of the synthesized beamwidth).

\begin{table}[htb]
\begin{tabular}{l|c}
\hline
Time (\hbox{MJD}, @2.0~GHz) & 58648.05071771 \\
R.A. (J2000) & 4h20m18.13s \\
Declination (J2000) & $+$73d42m24.3s \\
R.A. (J2000, deg) & 65.07552 \\
Declination (J2000, deg) & 73.70674 \\
Centroid ellipse (\arcsec, \arcsec, \arcdeg) & 0.8\arcsec, 0.4\arcsec, 67 \\
S/N ratio$_{\rm{image}}$ & 8.27 \\
DM$_{\rm{obs}}$ ($\dmunits$) & 959.2$\pm5$ \\ 
DM$_{\rm{MW}}$ ($\dmunits$) & 83.5 \\
Peak flux density (mJy) & 124$\pm$14 \\
Fluence (Jy~ms) & 0.62$\pm$0.07 \\
Deep limit ($\mu$Jy/beam) & $<$11 \\ \hline
\end{tabular}
\caption{Measured properties of \frbname\ with $1\sigma$\ errors. The centroid ellipse is defined with the major and minor axes and orientation (east of north). Deep limit refers to the flux density limit on 1.4~GHz radio counterparts in a deep image of the FRB field. The Milky Way DM estimate is calculated from \citet{2002astro.ph..7156C}. \label{tab:prop}}
\end{table}

\subsection{CHIME/FRB Limits}
The CHIME/FRB system, operating in its commissioning phase, has observed the sky position of \frbname\ for a total of 153~hours during the interval from 2018 August~28 to 2019 September~30. The large exposure is due to the circumpolar nature of the source and is split between 88~hours for the upper transit and 65~hours for the lower transit. The average duration of the upper and lower transits is 17 and 13 min, respectively, during which the source is within the FWHM region of the synthesized beams at 600 MHz. We searched through all low-significance events that were detected by the CHIME/FRB system in the above-mentioned observing time. No significant event or excess event rate was found to be consistent with the location and DM of \frbname, so there is no evidence for repetition from this FRB.

\begin{figure}[tb]
\begin{center}
\includegraphics[width=\columnwidth]{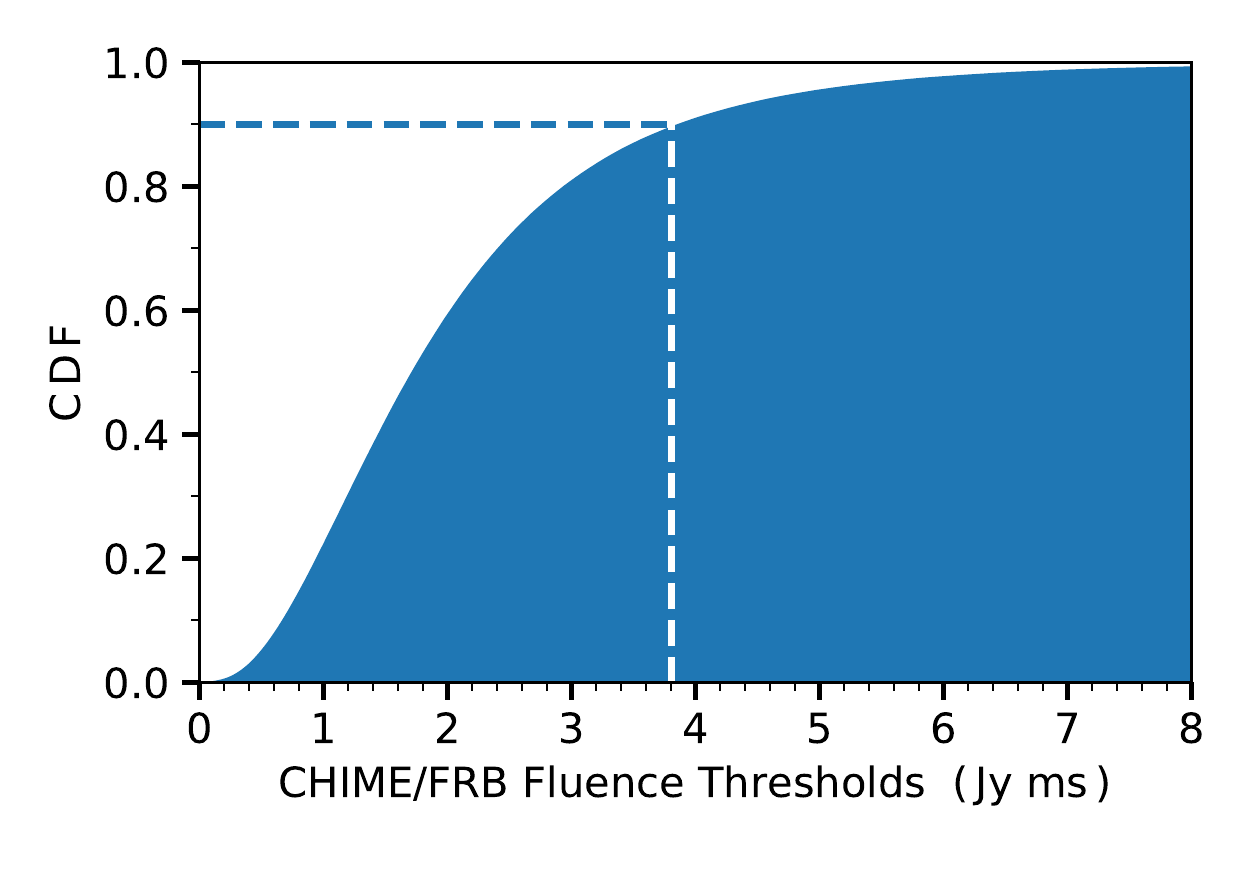}
\caption{Cumulative distribution of fluence detection thresholds for the CHIME/FRB instrument. Note that the FRB candidate is circumpolar and thus transits the CHIME FoV twice a day; thresholds shown here are valid for the upper transit, whereas the lower transit is a factor of ${\sim}$4 less sensitive. Dashed lines indicate the 90\% completeness level at 3.8 Jy ms. For comparison, the VLA fluence limit is 0.5 Jy ms (8$\sigma$ in 5 ms at 1.4 GHz).
\label{fig:chime}}
\end{center}
\end{figure}

To determine CHIME/FRB sensitivity to \frbname, we follow the methods detailed in \citet{2019ApJ...882L..18J}. The sensitivity of CHIME/FRB varies with observing epoch, position along transit, and a burst spectral shape. We used a Monte Carlo simulation with 10$^6$\ realizations to generate fluence thresholds for different detection scenarios within the quoted exposure. These simulations define a set of relative sensitivities, which are tied to a flux density scale using beam-formed, bandpass-corrected observations. As a reference, we use a burst from FRB 180814.J0422+73 detected on 2018 November~11 with a S/N ratio of 9.8$\sigma$, fluence of $2.3\pm0.8$\ Jy~ms, and a Gaussian spectral fit with center frequency of 524~MHz and FWHM of 72~MHz. Figure~\ref{fig:chime} shows the fluence threshold distribution is 90\% complete at 3.8~Jy~ms. The distribution is valid for the upper, more-sensitive transit; we estimate the lower transit to be approximately a factor of 4 less sensitive \citep{2019Natur.566..235C}.

\begin{figure*}[tb]
\begin{center}
\includegraphics[width=\textwidth]{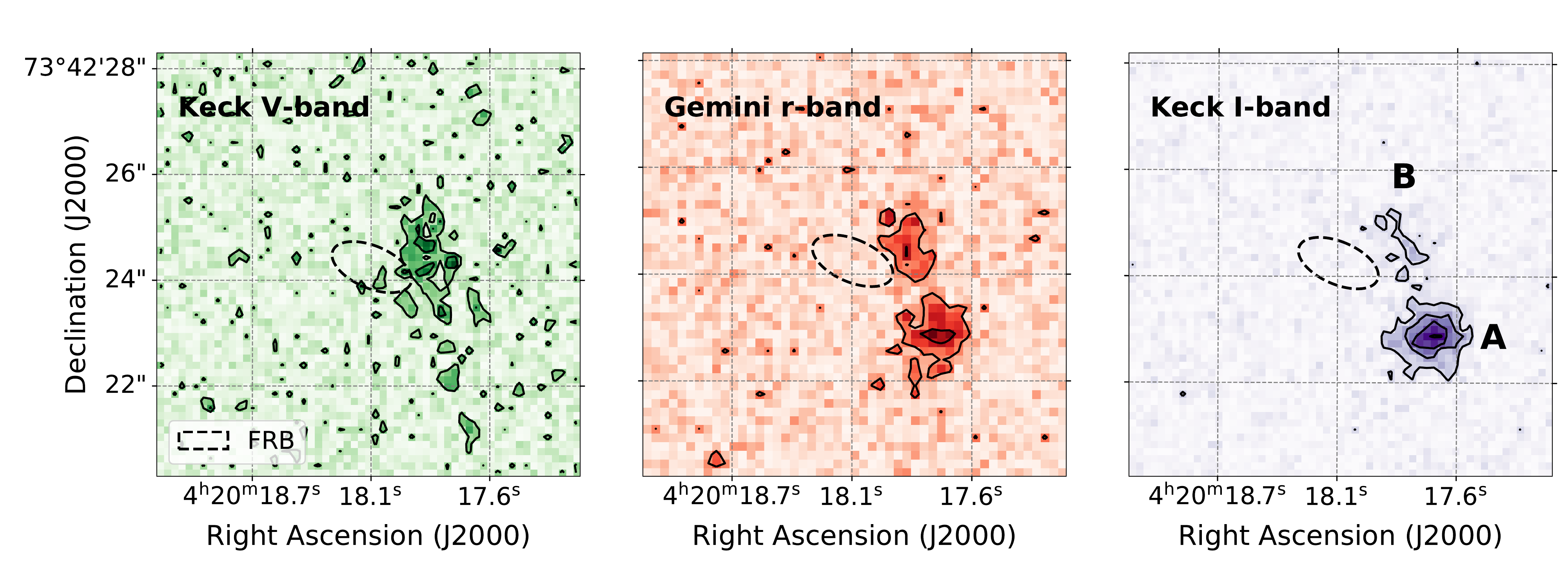}
\caption{Cut-out images from Keck/LRIS and Gemini/GMOS centered on the candidate FRB. The dashed line shows the $1\sigma$\ radio centroid region. Source A (brighter, to south) is red with brightest flux in the $I$~band. Source B (fainter, to north) is bluer with colors indicative of star formation.}
\label{fig:imageo}
\end{center}
\end{figure*}

\begin{figure*}[tb]
\begin{center}
\plottwo{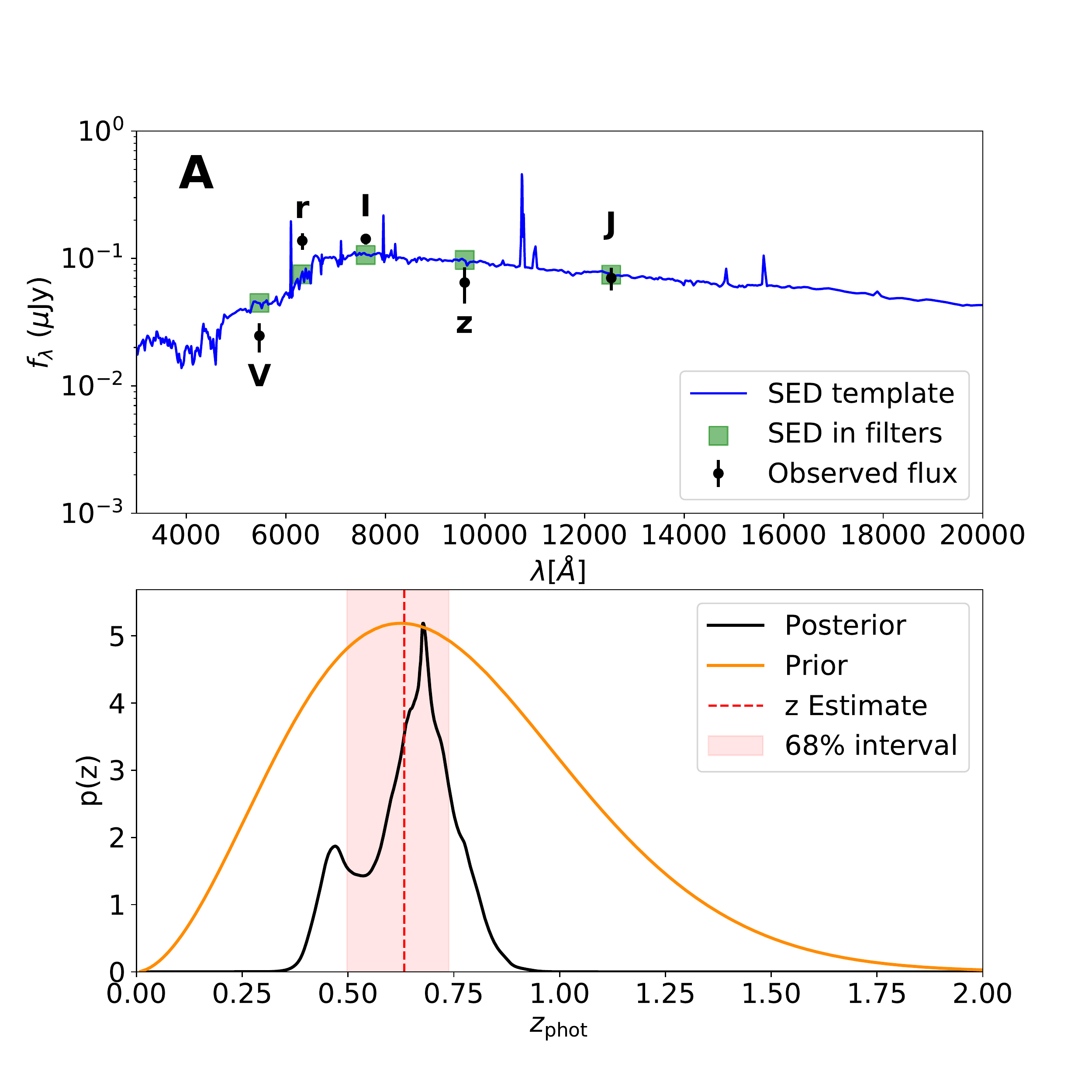}{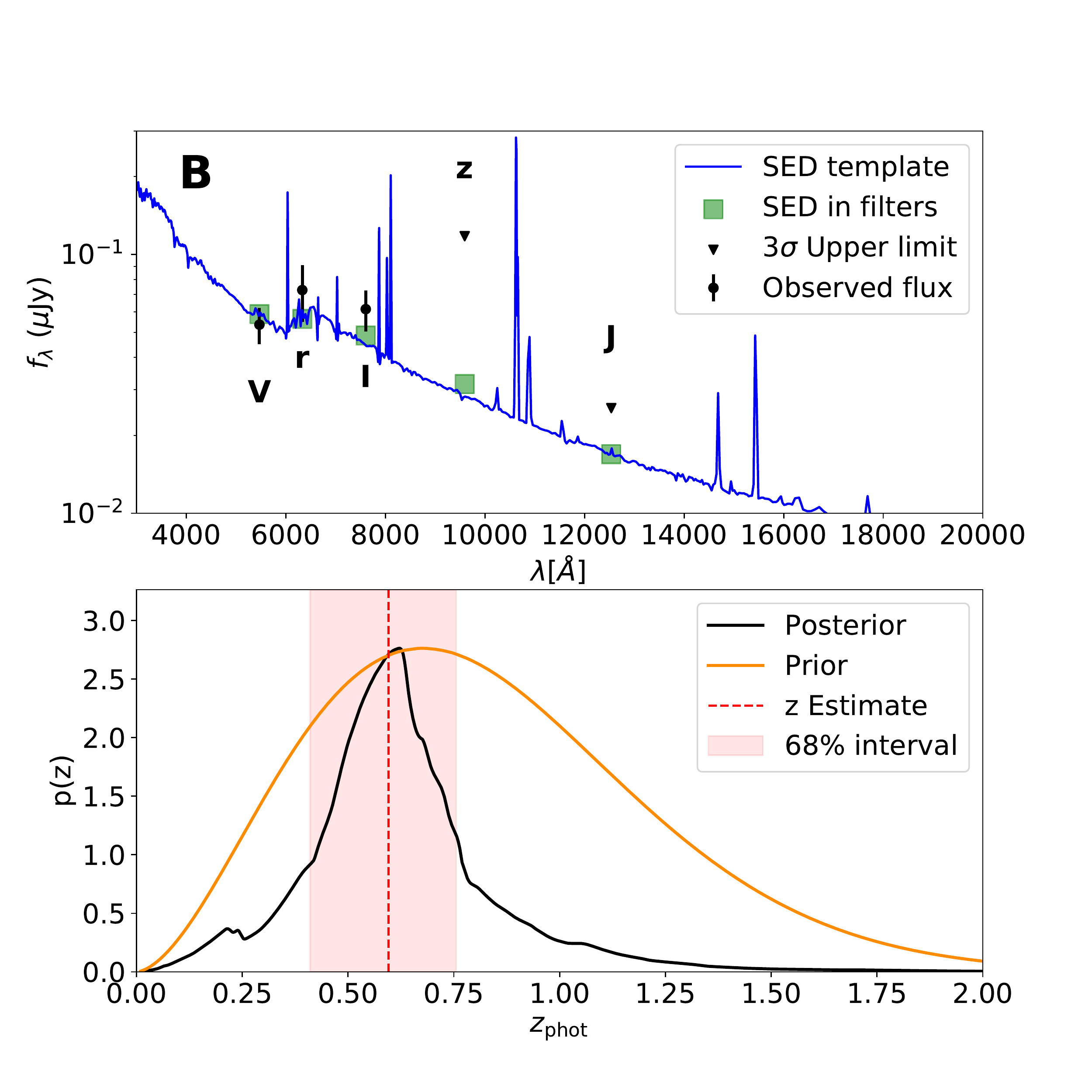}
\caption{(Top left) The photometric measurements of source~A with best-fit model in blue. SED in filters shows the best-fit template fluxes in each filter, with the black points showing the measured flux in the filters. (Bottom left) The redshift posterior for source~A as estimated by {\sc eazy}. The red dashed line shows the expectation value of the redshift over the posterior. With the pink shaded region marking the 16-84th percentile range. The stated $\sigma$ associated with $z_{phot}$ is half of the difference between the upper and lower limits shown above.
(Right) Same left panels, but for source~B.
%
\label{fig:photoz}}
\end{center}
\end{figure*}

\subsection{Optical Associations}
\label{sec:optical}
We considered the significance of this candidate high enough to trigger observations designed to find an optical counterpart. On UT 2019 July~2, we observed the field surrounding \frbname\ with the Gemini Multi-Object Spectrograph (GMOS) on the Gemini-N telescope. We obtained a series of $8\times300$~s image exposures in the $r$-band.  These data were reduced with standard procedures using the Gemini's {\sc pyraf} package,
and the images were registered using Pan-STARRS DR1 astrometric standards \citep{Flewelling16}. We performed photometry on these images using {\tt DoPhot} \citep{Schechter93} and calibrated the image using Pan-STARRS $r$-band calibrators.

On UT 2019 September 25, we obtained a series of 4$\times$600~s images with the Low Resolution Imaging Spectrograph (LRIS) on the Keck~I telescope in $V$ and $I$ bands. These data were reduced using a custom-built pipeline used for transient searches and based on the {\tt photpipe} imaging and reduction package \citep{Rest05}. Following standard procedures, we removed bias and flattened our images using bias and dome flat-field exposures obtained on the same night and in the same instrumental configuration. We registered the images using Pan-STARRS astrometric standards and combined the individual exposures with {\tt SWarp} \citep{Bertin02}. We performed point-spread function photometry on the final stacked images with {\tt DoPhot} and calibrated these data using Pan-STARRS $grizy$ calibrators transformed to $VI$ using the bandpass transformations described in \citet{Tonry12}.  

On UT 2019 November 26, we obtained an additional set of $18\times200$~s $z$-band images of the FRB field with the Alhambra Faint Object Spectrograph and Camera (ALFOSC) on the Nordic Optical Telescope (NOT). The images were processed with standard procedures and astrometrically-calibrated to the {\it Gaia}-DR2 reference frame.

On UT 2020 March 09, we also obtained a set of $4 \times 300$~s images (each one coming from $5 \times 60$~s co-adds) in the near-infrared J-band using the Near InfraRed Imager and spectrograph \citep[NIRI;][]{niri} on the Gemini-N telescope. The images were reduced with standard procedures using the {\sc dragons}\footnote{\url{https://dragons.readthedocs.io}} package and were astrometrically-calibrated to the {\it Gaia}-DR2 reference frame. A photometric calibration was derived using 2MASS sources in the image.

Figure~\ref{fig:imageo} shows the $VrI$ images centered on the radio localization of the candidate FRB. All optical images were registered in the Pan-STARRS DR1 astrometric frame, and so the uncertainty in their relative alignment is given by the precision of the original alignment solutions. We estimate a registration precision of $\approx0.06\arcsec$\ (1$\sigma$) for each image. 

There are two optical sources that are plausibly associated with the radio source. The brighter is J042017.85+734222.8, referred to as Source~A, and approximately 1\arcsec\ north of that is J042017.86+734224.5, referred to as source~B. The $1\sigma$ radio localization region overlaps with source~B, but the $2\sigma$ (90\% confidence interval) radio localization region overlaps with source~A. We consider both sources as potentially associated with the event. Final $VrIzJ$ photometry of the candidate FRB hosts was obtained using a 1\arcsec\ aperture centered at the locations described in \autoref{tab:sources} and corrected for Galactic extinction.

With the photometry of the galaxies as inputs, we have used the software package {\sc Eazy} \citep{Brammer2008} to estimate photometric redshifts for the two sources closest to \frbname. We find $z_{\rm phot} = \zaphot$ (68\% confidence interval) for source~A, and $z_{\rm phot} = \zbphot$ (68\% confidence interval) for source~B. 
Figure~\ref{fig:photoz} shows the redshift posterior distributions for sources~A and~B and their best-fitting templates. The best-fitting template for source~A is a relatively quiescent galaxy with weak emission features whereas \hbox{source~B}, which exhibits a bluer color, is best-fit with a star-forming template. The SED templates that were fitted to the data, agree well with the color difference that we observe in the source. With testing multiple sets of SED templates, we consistently find source~A to be quiescent, and similar in shape to what is shown above, as well as source~B consistently being fit to star-forming, bluer templates. We also note that the $u-r$ restframe colors from the CIGALE analysis detailed below,  are consistent with the {\sc eazy} outputs.

On UT 2019 September 29, we obtained a series of long-slit spectra ($1''$ wide) of source A and B with LRIS configured to cover wavelengths $\lambda \approx 3200 - 6800$\AA\ with the blue camera and its 300/5000 grism and $\lambda \approx 6720-9090$\AA\ with the red camera using the 831/8200 grating. These data were reduced and calibrated with the {\sc PypeIt} software package \citep{pypeit}. While we detect a very faint trace of continuum emission from source~A, there is no obvious emission or absorption feature to establish a spectroscopic redshift. This is consistent with it being an early-type galaxy with low or negligible star-formation and correspondingly weak nebular emission.  We did not identify any significant flux
from source~B.

To roughly estimate the stellar mass and rest-frame color of each candidate host galaxy, we performed a spectral energy distribution (SED) analysis of the measured photometry (Table~\ref{tab:sources}). This analysis, using the CIGALE software package \citep{Noll:2009aa}, also requires the source redshift; we adopted the posterior-weighted photometric redshift from the {\sc eazy} analysis (Table~\ref{tab:sources}). For the SEDs constructed by CIGALE, we adopt a delayed-exponential star-formation history model  with no late burst population, a Chabrier initial mass function, and the \cite{Calzetti:2000lr} dust extinction model. Because of the applied extinction corrections, changes in these assumptions would produce similar results.  Consistent with the {\sc eazy} analysis, the best-fitting SEDs were quiescent for source~A and star-forming for source~B. In Table~\ref{tab:sources}, we report estimates for the stellar mass and rest-frame $u-r$ color with the latter reflective of the inferred star-forming properties of each galaxy.  We caution that the stellar mass, especially, bears great uncertainty due to the uncertain redshifts of each source.

\begin{deluxetable*}{cccccc}
\tablewidth{0pc}
\tablecaption{Optical Candidates\label{tab:sources}}
\tabletypesize{\footnotesize}
\tablehead{ & & \multicolumn{2}{c}{Source A} & \multicolumn{2}{c}{Source B} \\ 
\colhead{Quantity}  
& \colhead{Unit} & \colhead{Value} & \colhead{Error} 
& \colhead{Value} & \colhead{Error} 
} 
\startdata 
RA (J2000) & deg& 65.07380& 0.00005& 65.0745& 0.0001\\ 
Dec (J2000) & deg& 73.70636& 0.00005& 73.7068& 0.0001\\ 
$V$ & mag& 25.42& 0.25& 24.58& 0.16\\ 
$r$ & mag& 23.25& 0.15& 23.94& 0.24\\ 
$I$ & mag& 22.83& 0.10& 23.74& 0.18\\ 
$z$ & mag& 23.18& 0.30& 22.53& 999.\\ 
$J$ & mag& 22.56& 0.20& 23.66& 999.\\ 
$z_{\rm phot}$ & & 0.63& 0.12& 0.60& 0.17\\ 
$\log_{10} \, M_*$ & $M_\odot$& 9.6 && 8.8 &\\ 
$u-r$ & mag & 2.1 && 0.8 &\\ 
\hline 
\enddata 
\tablecomments{This AB photometry has been corrected for Galactic extinction. 
A 999.9 value for photometric error indicates a $3\sigma$ upper limit. 
Estimates for $M^*$ and $u-r$ are based on the photometric redshift and bear great uncertainty.} 
\end{deluxetable*}

\section{Discussion}\label{sec:discuss}

\subsection{Joint Probability of Radio Candidate with Optical Association}
\label{sec:assoc}

The chance of randomly associating a point on the sky with a galaxy has previously been studied in the context of gamma-ray bursts. The chance association has an empirically-defined functional form parameterized by a association tolerance and survey depth \citep{2002AJ....123.1111B}\footnote{Eqs.~1-3 of \citet{2002AJ....123.1111B} are implemented in \url{https://github.com/FRBs/FRB}.}. Following the same approach we estimate chance association probabilities of $P_{\rm ch,A} =  7.1\times 10^{-2}$ and $P_{\rm ch,B} = 6.9\times 10^{-2}$ for the two galaxies to be unrelated to \frbname\ based on the $r$-band detections of Galaxy~A ($r=23.24$\,mag) and B~($r=23.93$\,mag). The maximum ``search radius'' $r_{\rm ch}$ used for these estimates take into account the half-light radii $R_{1/2}$ of the host candidates and the distance to the galaxy centroids ($d_{\rm gal,A} = 2.22\arcsec$ and $d_{\rm gal,B} = 1.07\arcsec$) as $r_{\rm ch} = \sqrt{d_{\rm gal}^2 + 4R_{1/2}^2}$. We also use the background flux as a limit on the presence of a galaxy below the detection limit of $r>25.1$\,mag. At this limit, the chance association probability of an unrelated galaxy to be located only within the error region of the FRB position is $P_{\rm ch,undet} =  1.08\times 10^{-2}$. The probability that either source~A and source~B are unrelated to the FRB is therefore small and the expectation for an even fainter host galaxy counterpart within the error region is even smaller.

We also used the methods of \citet{2017ApJ...849..162E} to calculate the chance association probabilities\footnote{Implemented in \url{https://github.com/KshitijAggarwal/casp} (Aggarwal et al, in prep).}. Using this approach, we estimate the chance coincidence probability of $P_{\rm ch,A} = 2.6\times 10^{-2}$ and $P_{\rm ch,B} = 2.7\times 10^{-2}$ for Galaxy~A and~\hbox{B}, respectively. Although, \citet{2017ApJ...849..162E} followed a similar procedure to \citet{2002AJ....123.1111B}, the estimates using their methods are smaller because they used a more recent estimate of $r$-band number counts of galaxies \citep{2016ApJ...827..108D} to calculate the number density of galaxies above any given limiting magnitude. We use the more conservative and more widely used estimates obtained using the formalism of \citet{2002AJ....123.1111B} for calculating the significance of the FRB candidate.

Under the assumption that an FRB should reside in a galaxy, we can use the host galaxy association to improve the confidence in the significance of the candidate event. The radio signal alone has been characterized by a SNR of 8.27 and a FAR of $4\times10^{-3}\ \rm{hr}^{-1}$. If we assume that false positives are randomly distributed in the field, then the association of the radio source to a host galaxy improves the confidence in the significance of the FRB candidate as $\rm{FAR}_{\rm{assoc}}=\rm{FAR} \cdot P_{\rm ch,det}$. According to this relation, we find $\rm{FAR}_{\rm{assoc}}=3\times10^{-4}\ \rm{hr}^{-1}$.

Given that the association of a false positive with a host galaxy is unlikely, we conclude that the FRB candidate is an astrophysical event. We used the Transient Name Server\footnote{See \url{https://wis-tns.weizmann.ac.il/}.} to name the event \frbname. \add{This naming convention is consistent with a new standard developed by several groups in the FRB community. The common convention used prior to this change is suitable as a shorthand and is "FRB 190614".}

\subsection{FRB Host Galaxy and DM}\label{sec:host}

The identification of a specific FRB host galaxy can be critical for both estimating the likely host DM contribution to the total observed DM, and for identifying trends in FRB host galaxy types and environments, which can in turn help discriminate between FRB origin models. \frbname\ is offset from optical counterpart(s), the components of which appear to be galaxies that differ in their mass, color, and type.

Given the total extent of the optical counterparts and the color differences,
it is most likely that they are two distinct galaxies. However, the photometric redshifts of the two galaxies are consistent with each other and they are close in projection. Assuming a distance of $z=0.6$ (6.8\,kpc/$''$), the galaxy centers are separated by 13.6\,kpc in projection. Assuming that the two galaxies are located at the same redshift, they are likely an interacting pair, in which source B may be a star-forming dwarf satellite of source A. If instead we do not assume they are interacting,
this projected separation can occurs by chance in galaxies of this magnitude about 10\% of the time.
In this case, one galaxy is the foreground object to the other.

While the optical data do not directly indicate which galaxy might be in the foreground, an analysis of dispersion does provide some hints. The net observed DM is a sum of contributions from: the Milky Way's interstellar medium and its halo, diffuse contributions from IGM plasma, any intervening galaxies and their related circumgalactic media, any cluster plasma, and any host galaxy or FRB-engine contribution \citep{2020arXiv200513157S}
. Any distant contribution is cosmologically redshifted, causing the rest-frame DM contribution to scale by $(1+z)^{-1}$ \citep[\eg][]{ymw17}.

Regarding local contributors to DM, for contribution from the interstellar medium of the Milky Way, we adopt the value of 83.5~$\dmunits$ predicted by \citet{2002astro.ph..7156C}. 
Generally contributions from Milky Way's ionized halo are taken to  30-80~$\dmunits$; here we adopt the model of \citet{prochaska+zheng2019}, which predicts a halo contribution of 64~$\dmunits$ for this sightline. Given these local contributions, we arrive at a representative extragalactic contribution of ${\rm DM_x} =812\pm25~\dmunits$, which encompasses all non-local contributions.

\begin{figure}
    \centering
    \includegraphics[width=\columnwidth,trim=0mm 0mm 13mm 3mm,clip]{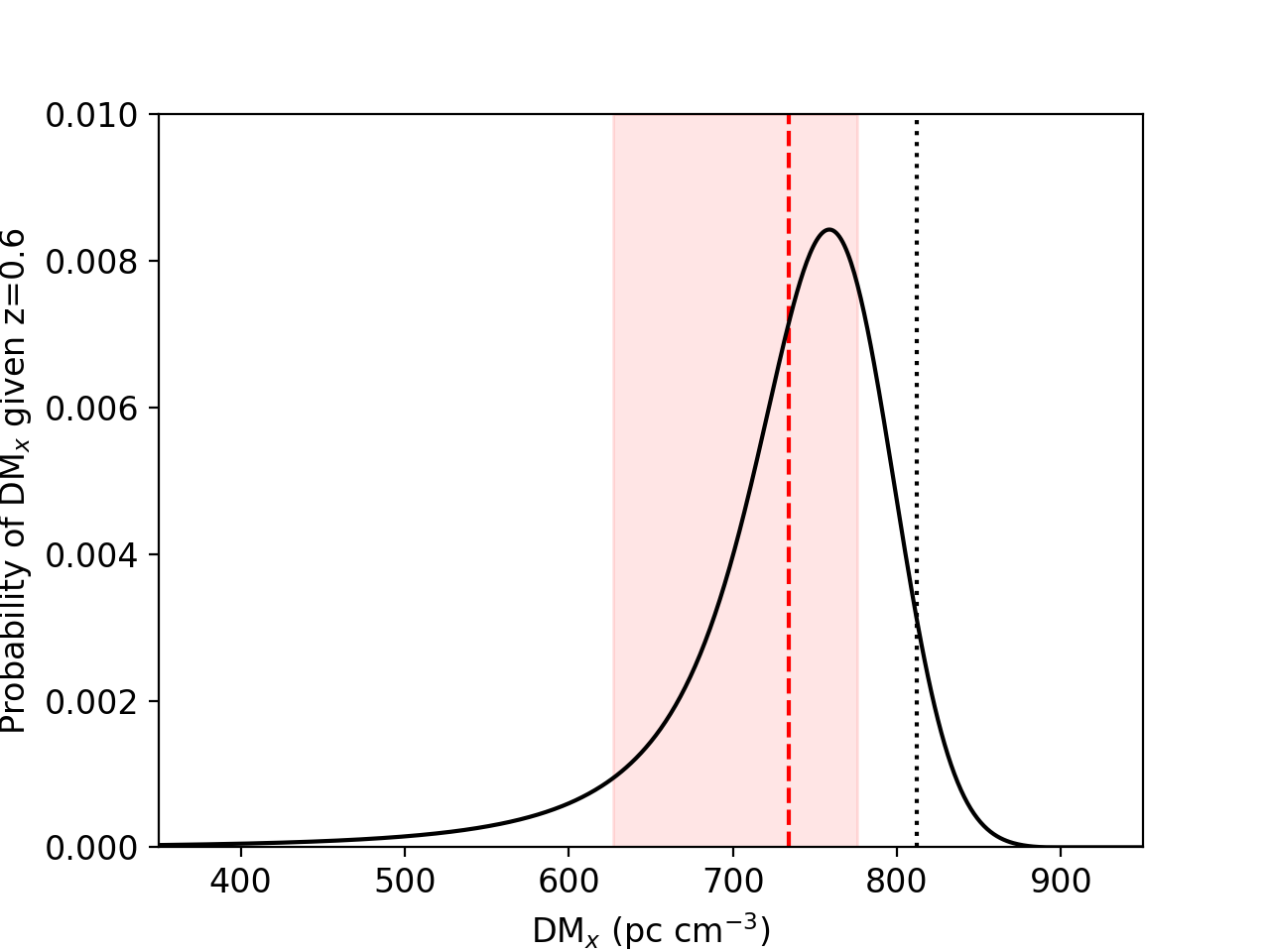}
    \caption{The range of extragalactic DM contributions (predominantly from the IGM and galaxy group halos) predicted by the model of \citet{prochaska+zheng2019}. The mean and 68\% range of the distribution is shown in red. The nominal value of DM$_{\rm x}=812\,\dmunits$ inferred for \frbname\ in Section \ref{sec:host} is shown here as a dotted black line.
    }
    \label{fig:dmdist}
\end{figure}

We can \add{use the scaling of DM with redshift \citep[known as the Macquart relation;][]{2020arXiv200513161M}} to estimate a \remove{DM-based} maximum possible redshift for our FRB. \add{To do this,} we attribute all of ${\rm DM_x}$ to an IGM that is devoid of cluster and galaxy group halos. There are various models that predict the ionization and elemental make-up of the IGM as a function of redshift; most of these provide results in the same range \citep[\eg][]{ymw17,2019ApJ...886..135P,prochaska+zheng2019}, predicting maximum redshifts in the $z=1.1 - 1.3$\ range. These values are well beyond the photometric redshifts we measured for both candidate hosts, implying that there are other significant contributors to DM than the IGM for these sight lines.

Figure~\ref{fig:dmdist} shows the expected probability distribution of non-local DM components using the model of \citet{prochaska+zheng2019}. The distribution assume an FRB located at $z=0.6$ and uses \add{a} parameterized model for halos from individual galaxies, groups, and clusters\footnote{Model implemented in \citet{j_xavier_prochaska_2019_3403651} with cosmological parameters described in \citet{2016A&A...594A..13P}}. In this formulation, the mean and 68\% range of this distribution gives ${\rm DM}=734^{+42}_{-107}\add{\,\dmunits}$. The estimated {$\rm DM_x$} for \frbname\ is not consistent with the predicted 68\% range at this redshift. \add{Note that the peak $DM_{\rm x}$ probability lies at a higher value than the mean, however even if we use that as a reference point for indicating potential host contributions, this minimal difference does not change the conclusions or analysis presented below.}

There are a few uncertainties in the comparison of measured to expected DM. First, this estimate ignores the host DM, both from the galaxy halo and interstellar medium. Any such component would push the predicted distribution to higher DM, making it more consistent with the observed value. However, this correction term is diminished from the rest frame value by $1+z_{\rm c}$, where $z_{\rm c}$ is the host redshift. For $z=0.6$, a rest frame host contribution would be roughly $57~\dmunits$\ to make the 68\% interval of the prediction consistent with the observed value. Second, the DM estimate for the Milky Way tends to be underestimated because the models do not include all small-scale contributions (\eg\ H$\alpha$ features) of roughly $\sim$10-20\% of the total DM column. Given this, the predicted DM is marginally consistent with expectations at the best-fit photometric redshift. This tension could be resolved if the FRB host galaxy is more distant than the second, non-host galaxy of the pair.

\add{While the above constraints don't provide conclusions about which of the two sources the FRB is likely associated with (either galaxy or an interacting pair could provide sufficient host DM contribution), 
it is worth making the simple note that the FRB's location appears to be closer in projection to source B, which we found to have a likely bluer stellar component and potentially more star-formation. Potential links with star-forming regions in galaxies have been noted for several past FRBs, particularly the localized repeating FRBs \citep[\eg][]{2020Natur.577..190M,OPT,2019Sci...366..231P}.}

\section{Conclusions}
We present the discovery of \frbname\ with VLA/\rf, the first FRB discovered blindly via interferometric imaging. The \rf\ system has a relatively high sensitivity and localization precision, which makes it possible to identify distant FRBs and associate them to host galaxies. We describe how the use of interferometric images enable simultaneous noise estimates that are more robust than the traditional time-local noise estimates. The radio event significance is low, but we argue that the nature of the radio measurement, considered with its association to a pair of host galaxies, is consistent with an astrophysical origin.

\frbname\ has the highest DM of any well-localized FRB ( DM$_x\approx812~\dmunits$) and is likely associated with a pair of host galaxies that are among the most distant hosts identified ($z\sim0.6$). At this distance, the burst energy is $\sim10^{31}$\ erg Hz$^{-1}$; the fluence, distance, and energy make this a faint version of the population typically seen by the Parkes Observatory \citep{2018Natur.562..386S}.

The DM is somewhat larger than predicted at the distance of the host galaxies, which implies a modest contribution from the FRB environment or intervening galaxy. The two associated galaxies differ in their colors and stellar masses, which implies different environments for the FRB. However, they are broadly consistent with Milky Way-like stellar masses and star formation rates, as has been identified in other FRB associations \citep{2020arXiv200513160B}.

The \rf\ system continues to commensally search for FRBs and other fast transients during VLA observations. In the future, the system will transition to a community service mode, in which real-time alerts are distributed automatically.

\acknowledgments

We thank the CHIME collaboration for supporting the analysis for this FRB and the NRAO for supporting \rf\ development and operations. We recognize and acknowledge the very significant cultural role and reverence that the summit of Maunakea has always had within the indigenous Hawaiian community. We are most fortunate to have the opportunity to conduct observations from this mountain.

CJL acknowledges support under NSF grant 2022546. BZ acknowledges the support of Frank and Peggy Taplin Membership Fund. On behalf of the F$^4$\ team, JXP, AM, NT, KEH, and SS acknowledge support under NSF grants AST-1911140 and AST-1910471. SBS and KA acknowledge support by NSF grant 1714897. SBS is a CIFAR Azrieli Global Scholar in the Gravity and the Extreme Universe program.
NT acknowledges support by FONDECYT grant 11191217. The NANOGrav project receives support from National Science Foundation (NSF) Physics Frontier Center award number 1430284. PC is supported by an FRQNT Doctoral Research Award.

The National Radio Astronomy Observatory is a facility of the National Science Foundation operated under cooperative agreement by Associated Universities, Inc. Part of this research was carried out at the Jet Propulsion Laboratory, California Institute of Technology, under a contract with the National Aeronautics and Space Administration. The Pan-STARRS1 Surveys (PS1) and the PS1 public science archive have been made possible through contributions by the Institute for Astronomy, the University of Hawaii, the Pan-STARRS Project Office, the Max-Planck Society and its participating institutes, the Max Planck Institute for Astronomy, Heidelberg and the Max Planck Institute for Extraterrestrial Physics, Garching, The Johns Hopkins University, Durham University, the University of Edinburgh, the Queen's University Belfast, the Harvard-Smithsonian Center for Astrophysics, the Las Cumbres Observatory Global Telescope Network Incorporated, the National Central University of Taiwan, the Space Telescope Science Institute, the National Aeronautics and Space Administration under Grant No. NNX08AR22G issued through the Planetary Science Division of the NASA Science Mission Directorate, the National Science Foundation Grant No. AST-1238877, the University of Maryland, Eotvos Lorand University (ELTE), the Los Alamos National Laboratory, and the Gordon and Betty Moore Foundation. Some/all of the data presented in this paper were obtained from the Mikulski Archive for Space Telescopes (MAST). STScI is operated by the Association of Universities for Research in Astronomy, Inc., under NASA contract NAS5-26555. Support for MAST for non-HST data is provided by the NASA Office of Space Science via grant NNX13AC07G and by other grants and contracts. This research has made use of NASA’s Astrophysics Data System.

Based on observations obtained at the international Gemini Observatory, a program of NSF’s OIR Lab, which is managed by the Association of Universities for Research in Astronomy (AURA) under a cooperative agreement with the National Science Foundation, on behalf of the Gemini Observatory partnership: the National Science Foundation (United States), National Research Council (Canada), Agencia Nacional de Investigaci\'{o}n y Desarrollo (Chile), Ministerio de Ciencia, Tecnolog\'{i}a e Innovaci\'{o}n (Argentina), Minist\'{e}rio da Ci\^{e}ncia, Tecnologia, Inova\c{c}\~{o}es e Comunica\c{c}\~{o}es (Brazil), and Korea Astronomy and Space Science Institute (Republic of Korea). The Gemini GMOS and NIRI data were obtained from programs GN-2019A-Q-107 and  GN-2020A-FT-201 (PI Spolaor) respectively, and it were processed using the Gemini's {\sc pyraf} and  {\sc dragons} packages\footnote{\url{https://www.gemini.edu/sciops/data-and-results/processing-software/}.} respectively.

Data were obtained at the W. M. Keck Observatory, which is operated as a scientific partnership among Caltech, the University of California, and the National Aeronautics and Space Administration (NASA). The Keck Observatory was made possible by the generous financial support of the W. M. Keck Foundation.

\vspace{5mm}
\facilities{EVLA, Pan-STARRS, MAST, Keck, Gemini, NOT}

\software{rfpipe \citep{2017ascl.soft10002L}, astropy \citep{astropy:2013,astropy:2018}, aegean \citep{2018PASA...35...11H}, {\sc dragons}, pypeit\citep{pypeit}, \citep{j_xavier_prochaska_2019_3403651}}

\bibliographystyle{aasjournal}
\bibliography{fasttrants.bib}

\appendix
\section{Robust Estimate of Event Significance with Interferometric Images}
\label{sec:std}

We describe the detection of an event with SNR estimate on the border between statistically significant and not. Therefore, we have to be very careful in assessing both it's power and the expected noise floor. A small change in the noise standard deviation estimate (and therefore the SNR) will dramatically change it's probability of chance occurrence. This type of error, although rarely considered, can push the detection threshold for FRBs and pulsars up by as much as 10\% in typical radio time-domain surveys. Correcting this type of error was shown to provide a substantial sensitivity improvement when applied to gravitational wave data in \cite{2019arXiv190805644Z}.

The SNR is derived from a detection score that is a particular linear combination of the data:

\begin{equation} \label{eq:S}
    S(\alpha,\delta) = \sum_{f,i,j,p}{G(i,j)V_{f,i,j,p}e^{i\phi_{f,i,j}(\alpha,\delta)}}
\end{equation}
Where $V_{f,i,j,p}$ are the visibilities of antennae $i,j$ at frequency $f$ and polarization $p$, $G_{i,j}$ are the empirically measured gains and $\phi$ are the calculated phases using the position $\alpha,\delta$. To a very good approximation, under the noise hypothesis and assuming no significant RFI, the score follows a Gaussian distribution. The tail of the Gaussian distribution is approximately proportional to 
\begin{equation}
     P(S>x)\propto e^{-x^2/2}
\end{equation}

For a Gaussian distribution, a 5\% change in the noise estimate translates to a factor of 30 in the chance occurrence probability of a tail event above a threshold of $S/N=8$. We therefore must have a noise standard deviation estimate that is good to $\approx 1\%$, which has less than a factor of two uncertainty in the chance occurrence probability. Empirically obtaining a 1\% estimate of the noise standard deviation requires of order $2\times 10^4$ independent measurements to average over. 

Obtaining such an accurate estimate is non-trivial in general. A common approach is to produce a time-series of detection scores at a single direction in the sky (either a single-dish beam or phased array of an interferometer), and computing a running standard deviation to use. However, in order to obtain $10^4$ independent samples few seconds of data are required, and slow gain fluctuations would typically bias the measurement. Our case was further complicated by the fact that the candidate FRB was discovered while the \rf\ system was turning on, so the number of recorded visibilities changes as a function of frequency/baseline/time. This precludes simple local noise estimates based on neighboring time or frequency samples.

We chose to use the image standard deviation as our noise estimate. To provide justification in using this estimate and to assess it's biases, we think of the phases in Eq. \ref{eq:S} as random, uncorrelated variables with uniform distribution in $[0,2\pi]$. From symmetry arguments, this standard deviation would depend only on the sum of the squared visibility amplitudes, or the total momentary power registered by all antennae \footnote{According to Parseval's theorem, the standard deviation of the image is exactly equal to the standard deviation estimate using the absolute value squared of the visibilities.}. Since the FRB contributes $\ll 1\%$ of the total power, it can be ignored. Therefore, the relative statistical error in the standard deviation is proportional to the inverse square root of the numbers of visibilities, which is smaller than 1\%\footnote{$N_{\rm vis} = N_{\rm bl} * N{\rm ch} * N_{\rm pol} * f_{\rm recorded} = 351 * 166 * 2 * 0.38 = 4.5e4$}.

We also note that if the empirically measured complex gains are accurate to 10\%, the SNR estimate we produce would be accurate to within 1\% of the optimal SNR with exact gains as the first order effect of gain errors (or FRB spectral shape) would cancel. See \cite{2019arXiv190805644Z} for a detailed computation on a similar problem for the case of detecting gravitational waves.

%
%

\end{document}